\documentclass{PoS}

\PoS{PoS(LAT2005)112}
\usepackage{here,epsfig,psfrag,cite}
\newcommand{\Tr}{\mbox{Tr}}
\newcommand{\beq}{\begin{equation}}
\newcommand{\eeq}{\end{equation}}
\newcommand{\la}{\langle}
\newcommand{\ra}{\rangle}

\title{A new efficient Cluster Algorithm for the Ising Model }

\ShortTitle{A new efficient cluster algorithm for the Ising model }

\author{\speaker{Matthias Nyfeler}, Michele Pepe, and Uwe-Jens Wiese\\
        Institute for Theoretical Physics, Sidlerstrasse 5, CH-3012 Bern, Switzerland\\
        E-mail: \email{nyfeler@itp.unibe.ch},  
	\email{pepe@itp.unibe.ch}, \email{wiese@itp.unibe.ch}}

\abstract{Using D-theory we construct a new efficient cluster algorithm
for the Ising model. The construction is very different from the standard
Swendsen-Wang algorithm and related to worm algorithms. With the new
algorithm we have measured the correlation function with high precision over
a surprisingly large number of orders of magnitude.}

\FullConference{XXIIIrd International Symposium on Lattice Field Theory\\
		 25-30 July 2005\\
		 Trinity College, Dublin, Ireland}

\begin{document}

\section{Reformulation of the Ising Model}

D-theory \cite{dtheory} is an alternative way of regularising field theories by 
dimensional reduction of systems involving 
discrete variables, that allows us to develop cluster
algorithms. One can also use this method to describe and simulate a 
simple spin system such
as the Ising model. 
Besides being interesting in its own right, we hope to gain some insights that
may be useful for the construction of cluster algorithms for gauge theories.

\begin{figure}[ht]
  \begin{center}
    \leavevmode
    \psfrag{b=eM}{$\beta=\epsilon M$}
    \psfrag{e}{$\epsilon$}
    \psfrag{T}{$T$}
    \psfrag{x}{$x$}  
    \includegraphics[height=3.5cm]{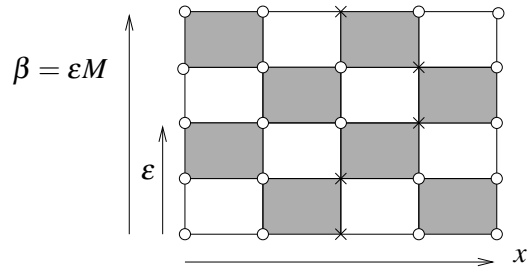}
  \end{center}
  \caption{Trotter decomposition
     in one dimension with a spin-configuration: crosses and circles represent 
     the two spin states.}
  \label{trotter}
\end{figure}

We consider the Ising model with the classical Hamilton function
${\cal H}[s]=-J\sum_{x,\mu}s_xs_{x+\hat{\mu}}$, 
and rewrite it as a quantum system with the Hamiltonian
$H=-J\sum_{x,\mu}\sigma_x^3\sigma_{x+\hat{\mu}}^3$. 
Due to the trace-structure of the partition function, $Z=\Tr\exp(-\beta H)$, we
can perform a unitary transformation rotating to a basis 
containing $\sigma^1$:
\beq H=-J\sum_{x,\mu}\sigma_x^1\sigma_{x+\hat{\mu}}^1.\eeq
We then perform a Trotter decomposition of the Hamiltonian. In one 
dimension this implies the following splitting into two parts,
\beq H=H_1+H_2,\ H_1=-J\sum_{x\ even}\sigma_x^1\sigma_{x+1}^1, \
H_2=-J\sum_{x\ odd}\sigma_x^1\sigma_{x+1}^1, \eeq
while in two dimensions one needs four parts
\beq H_1=\sum_{x\in (2m,n)}h_{x,\hat{1}},\ H_2=\sum_{x\in (m,2n)}h_{x,\hat{2}},\
  H_3=\sum_{x\in (2m+1,n)}h_{x,\hat{1}},\ H_4=\sum_{x\in
  (m,2n+1)}h_{x,\hat{2}}.
\label{trotter2d}\eeq
The partition function can then be turned into a path integral with $M$
time steps of width $\epsilon$ ($\beta=\epsilon M$). The Euclidean time serves as
an additional dimension whose extent $\beta$ determines 
the inverse temperature. For example in the 1-dimensional Ising model, the Trotter
decomposition of the partition function takes the form
\beq Z=\Tr\exp(-\epsilon H_1-\epsilon H_2)^M 
=\Tr\exp(-\epsilon H_1)^M\exp(-\epsilon H_2)^M .\eeq
In this way we obtain a checkerboard decomposition where the
action resides on the shaded plaquettes (see figure \ref{trotter}).

In general the Trotter decomposition is affected by an order $\epsilon^2$ error.
However, since in the Ising model $H_1$ and
$H_2$ commute, it is an exact rewriting of the partition function
and even with the minimal number 
of time-slices one obtains the continuous time limit.

\section{Construction of the Algorithm}

\begin{figure}[ht]
 \begin{minipage}[t]{0.5\linewidth}
  \begin{center}
  \leavevmode
    \includegraphics[height=0.9cm]{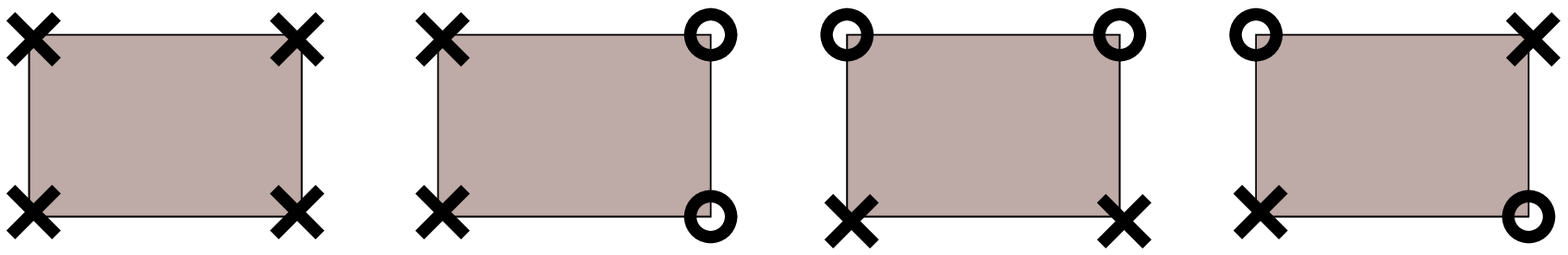}
    \caption{Allowed plaquettes.}
    \label{plaquettes}
  \end{center}
 \end{minipage}
 \begin{minipage}[t]{0.5\linewidth}
  \begin{center}
    \leavevmode
    \psfrag{A}{$A$}
    \psfrag{B1}{$B_1$}
    \psfrag{B2}{$B_2$}
    \includegraphics[height=0.9cm]{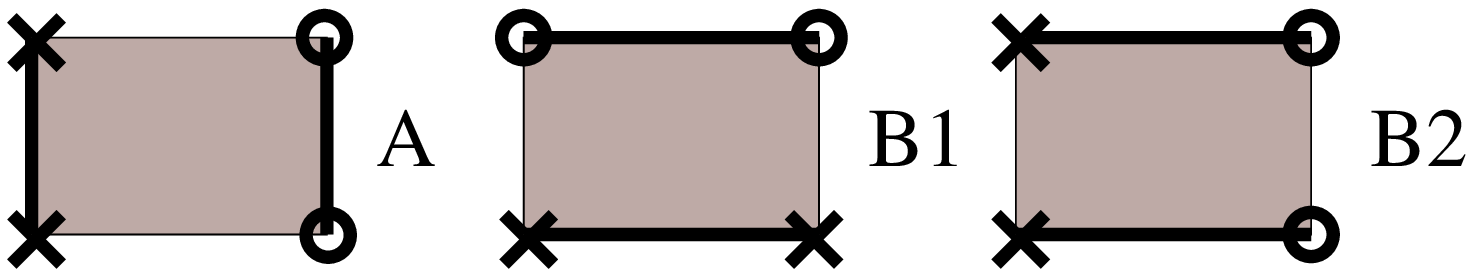}
    \caption{Cluster breakups.}
    \label{breakups}
  \end{center}
 \end{minipage}
\end{figure}

In our basis the transfer matrix of a single plaquette 
$T=\exp(-\epsilon \sigma_x^1\sigma_{x+1}^1)$ takes the form
\beq
{T=\small
\begin{array}{cc}
\begin{array}{cccc}
\uparrow\uparrow & \uparrow\downarrow & 
\downarrow\uparrow & \downarrow\downarrow
\end{array}
& \\
\left(\begin{array}{cccc}
c & 0 & 0 & s \\
0 & c & s & 0 \\
0 & s & c & 0 \\
s & 0 & 0 & c 
\end{array}\right) 
&
\begin{array}{c}
\uparrow\uparrow \\ \uparrow\downarrow \\
\downarrow\uparrow \\ \downarrow\downarrow \\
\end{array}
\end{array}} ,
\label{transfermatrix}\eeq
with $c=\cosh(\epsilon J)$ and $s=\sinh(\epsilon J)$. From this we
can see that there are only eight physically allowed
plaquette configurations (see figure \ref{plaquettes}).

In order to imply constraints on the relative orientation of the spins on a 
plaquette, we propose plaquette breakups that bind spins together. 
These constraints must be maintained in the cluster update. 
The breakups give rise to clusters and these clusters can only be
flipped as a whole. In this way, by construction, one indeed maintains the
constraints implied by the cluster breakups.
We choose to use the $A$, $B_1$, and $B_2$ breakups 
shown in figure \ref{breakups}.

\begin{figure}[ht]
  \begin{center}
    $\begin{array}{lll}
    \psfrag{A}{$A$}
    \psfrag{B1}{$B_1$}
    \psfrag{B2}{$B_2$}
    \includegraphics[height=3cm]{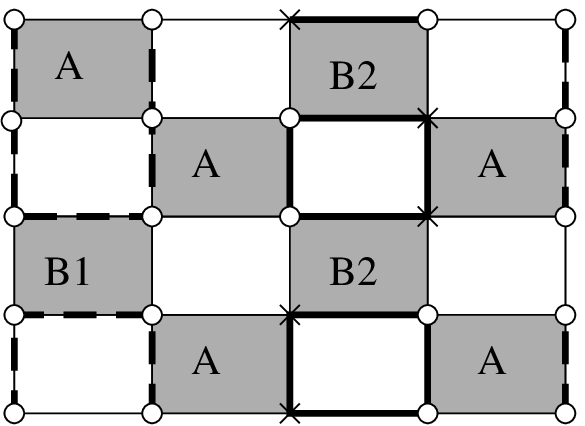}
    & \hspace{0.5cm} &
    \includegraphics[height=3cm]{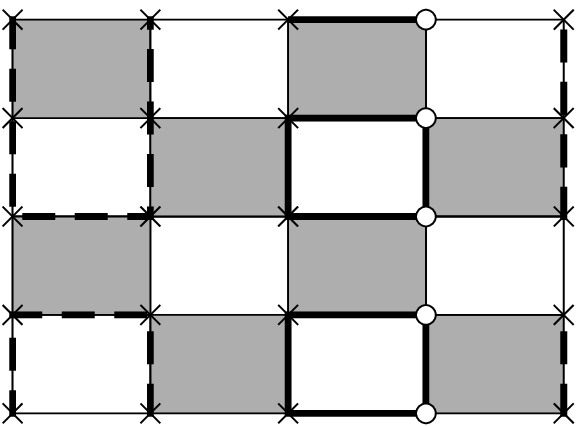}
    \end{array}$
  \end{center}
  \caption{A configuration with $A$, $B_1$, and $B_2$ breakups with a multi-cluster
  update.}
  \label{clusterflip}
\end{figure}

These breakups decompose the transfer matrix as follows:
\beq T=A\left( {\scriptsize \begin{array}{cccc}
1 & 0 & 0 & 0 \\
0 & 1 & 0 & 0 \\
0 & 0 & 1 & 0 \\
0 & 0 & 0 & 1 \\
\end{array}} \right) 
+B_1\left( {\scriptsize \begin{array}{cccc}
1 & 0 & 0 & 1 \\
0 & 0 & 0 & 0 \\
0 & 0 & 0 & 0 \\
1 & 0 & 0 & 1 \\
\end{array}} \right) 
+B_2\left( {\scriptsize \begin{array}{cccc}
0 & 0 & 0 & 0 \\
0 & 1 & 1 & 0 \\
0 & 1 & 1 & 0 \\
0 & 0 & 0 & 0 \\
\end{array}} \right) .\eeq
From eq.(\ref{transfermatrix}), one obtains $B_1=B_2=s$ and $A=c-s$.
Using these cluster breakups
we can now simulate the system. 
In figure \ref{clusterflip} we show an example of a cluster update in which the
dashed cluster has been flipped.

The energy density is related to the fraction $P$ of interaction 
plaquettes with a transition (spins opposite in time on a plaquette) by the 
equation
\beq E=\la H\ra=-J ( t+\frac{(1-t^2)}{t}\langle P\rangle ) ,\eeq
where $t=\tanh(\epsilon J)$.

\begin{figure}[ht]
  \begin{center}
    \leavevmode
    \includegraphics[width=5.5cm,angle=-90]{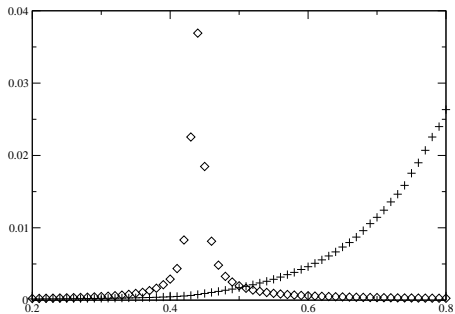}
  \end{center}
  \caption{Average cluster-size of the new algorithm (diamonds) compared 
  to the Swendsen-Wang algorithm (crosses) on a $100^2$ lattice.}
  \label{meancluster}
\end{figure}

Analysing the average cluster-size
of the algorithm in multi-cluster mode
shows that it is indeed very different from the Swendsen-Wang
algorithm \cite{Swendsen:1987ce}. In two dimensions, with the Trotter 
decomposition of eq.(\ref{trotter2d})
and the minimal number of four time-slices, we observe a peak
at the critical temperature, 
whereas Swendsen-Wang clusters still grow in the broken phase.
This is shown in figure \ref{meancluster}. 

\section{Measurement of the 2-Point Correlation Function}

\begin{figure}[H]
  \begin{center}
    $\begin{array}{ccc}
      \sigma^1={\small
        \begin{array}{cc}
          \begin{array}{cc} \uparrow & \downarrow \end{array}
            & \\
            \left(\begin{array}{cc} 0 & 1 \\ 1 & 0 \end{array}\right)
            & 
          \begin{array}{c} \uparrow \\ \downarrow \end{array}
        \end{array}}
        & \hspace{0.5cm} &
	\begin{array}{c}
	  \\
          \includegraphics[height=1.2cm]{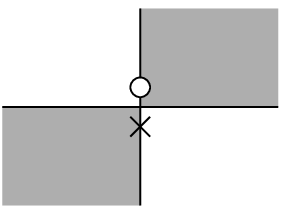}
	\end{array}
    \end{array}$
  \end{center}
  \caption{The action of the operator $\sigma^1$ on a site, seen as a violation
  of spin conservation.}
  \label{violation}
\end{figure}

The correlation function is defined in our basis 
as $\langle\sigma^1_0\sigma^1_x\rangle$. The
$\sigma^1$ matrices are off-diagonal and flip the spins they act on. Thus they
can be viewed as violations of spin conservation, i.e. two opposite spins reside
on the same site (see figure \ref{violation}).

\begin{figure}[ht]
  \begin{center}
    \psfrag{A}{$A$}
    \psfrag{B1}{$B_1$}
    \psfrag{B2}{$B_2$}
    \includegraphics[height=3cm]{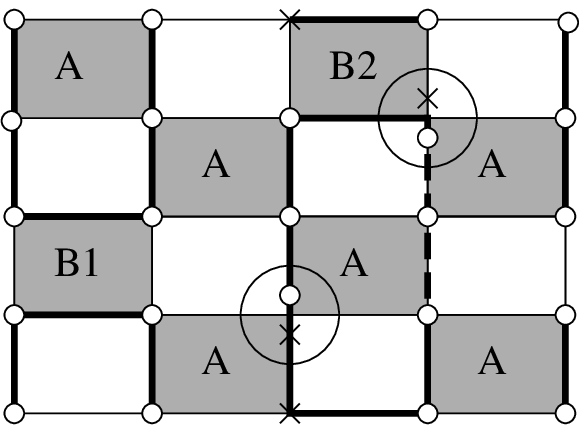}
    \hspace{0.5cm}
    \includegraphics[height=3cm]{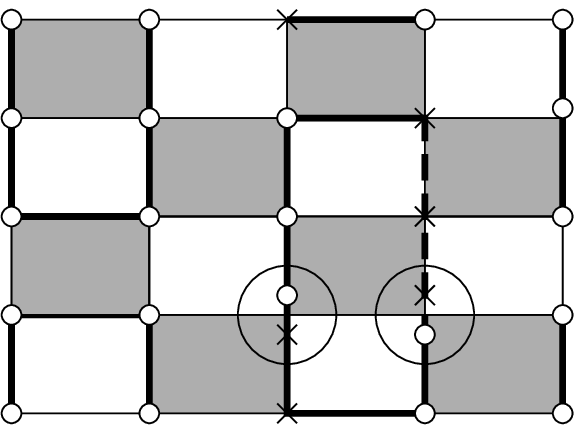}
  \end{center}
  \caption{Partially flipping a cluster with violations (circles).}
  \label{violationclusters}
\end{figure}

What we do is the following: we introduce two violations on a random site (where
they cancel each other), build the clusters, and flip them with probability
$p=\frac{1}{2}$, unless it is a cluster with one or two violations.
There we flip a fraction of the cluster in order to move
the violations to different positions.
For example in figure \ref{violationclusters}, the dashed part 
of a cluster with violations has been flipped.

When the violations are on the same site, we introduce them randomly 
somewhere else. By histogramming the spatial distance 
between the two violations,
we get a very accurate 
measurement of the correlation function. 
One then normalises the correlation function to 1 at zero distance.
In figure \ref{correlation} this is compared to the 
analytic result \cite{JimboMiwa}.

\section{Improved Estimator for the Susceptibility}

The susceptibility can be measured by summing over the
correlation function. This is already an improved estimator because one gets
only positive contributions.
One can also measure the average frequency of reaching spatial distance zero.
This corresponds to the inverse susceptibility.
In a large system at high temperature this is not too efficient yet and can be
further improved.

We project the clusters with violations to the spatial volume and
measure their overlap. This gives us the probability that two violations
are at zero spatial distance in the next cluster update and, again,
corresponds to measuring the inverse susceptibility.
In this way we gain statistics by a factor of the cluster-size 
squared $|C|^2$ with an effort proportional to the cluster-size $|C|$.

\section{Worm-Algorithm}

Instead of using clusters, we can --- with the same breakups --- move the two
violations locally with a worm-type Metropolis algorithm \cite{wormalgorithm}. 
\begin{figure}[ht]
  \begin{center}
    \leavevmode
    \psfrag{b=eM}{$\beta=\epsilon M$}
    \psfrag{e}{$\epsilon$}
    \psfrag{T}{$T$}
    \psfrag{x}{$x$}  
    \includegraphics[height=3cm]{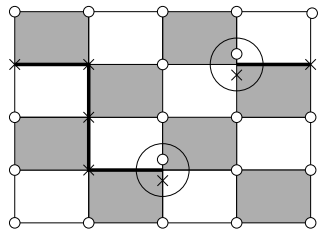}
  \end{center}
  \caption{Worm moving the violations and thus updating a given configuration.}
  \label{worm}
\end{figure}

Interestingly we can also measure n-point functions by introducing $n$ 
violations. These violations are moved in the same way as above.

\subsection{Snake-Algorithm --- Improved Correlation Function Measurement}

\begin{figure}[ht]
 \begin{minipage}[t]{0.48\linewidth}
   \begin{center}
    \includegraphics[width=5.5cm,angle=-90]{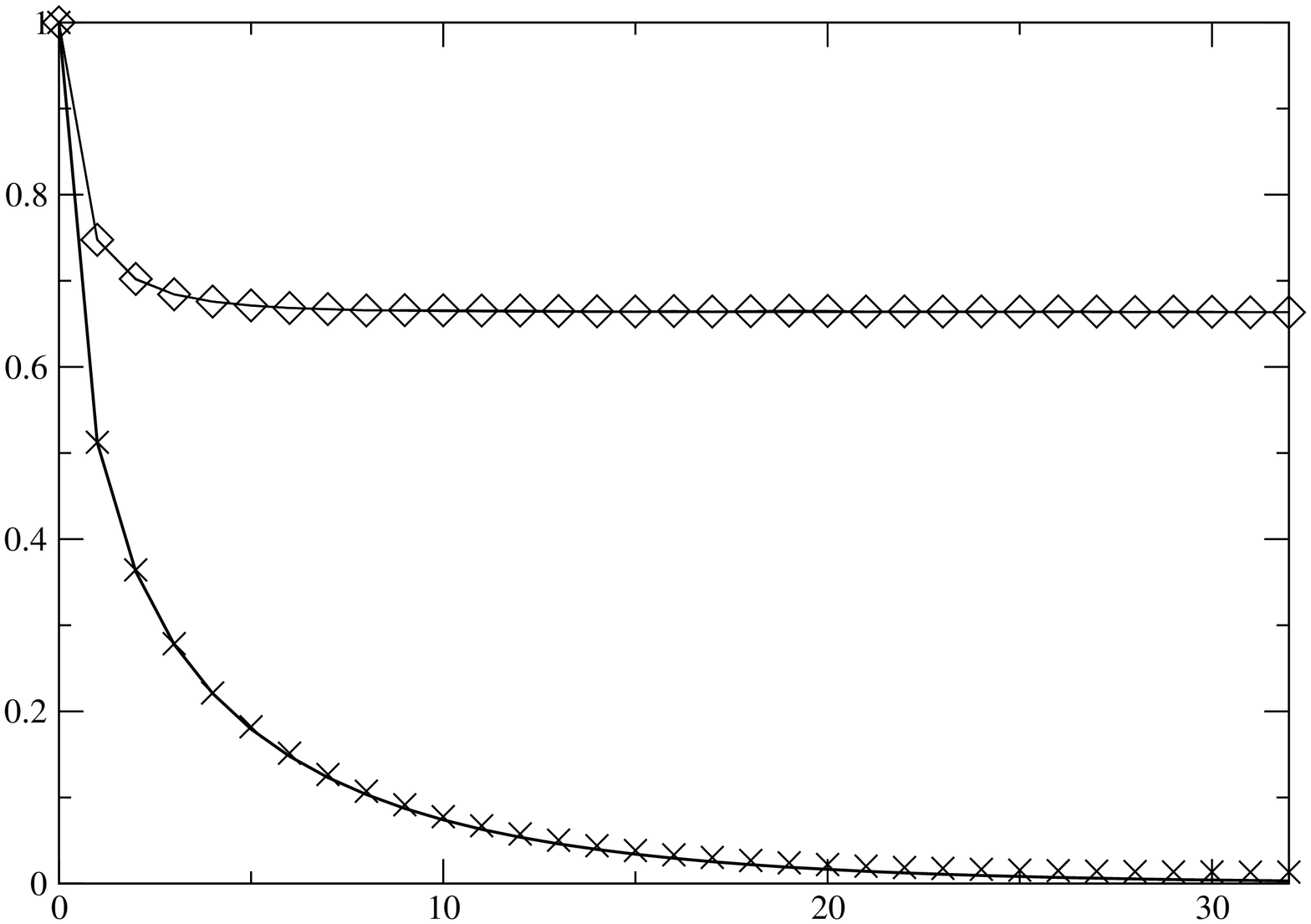}
  \end{center}
  \caption{Correlation function on a $64^2$-lattice at 
    $\beta=0.42$ (crosses) and $\beta=0.46$ (diamonds) compared to the analytic result.}  
  \label{correlation}
 \end{minipage}
 \hspace{0.04\linewidth}
 \begin{minipage}[t]{0.48\linewidth}
  \begin{center}
    \leavevmode
    \includegraphics[width=5.5cm, angle=-90]{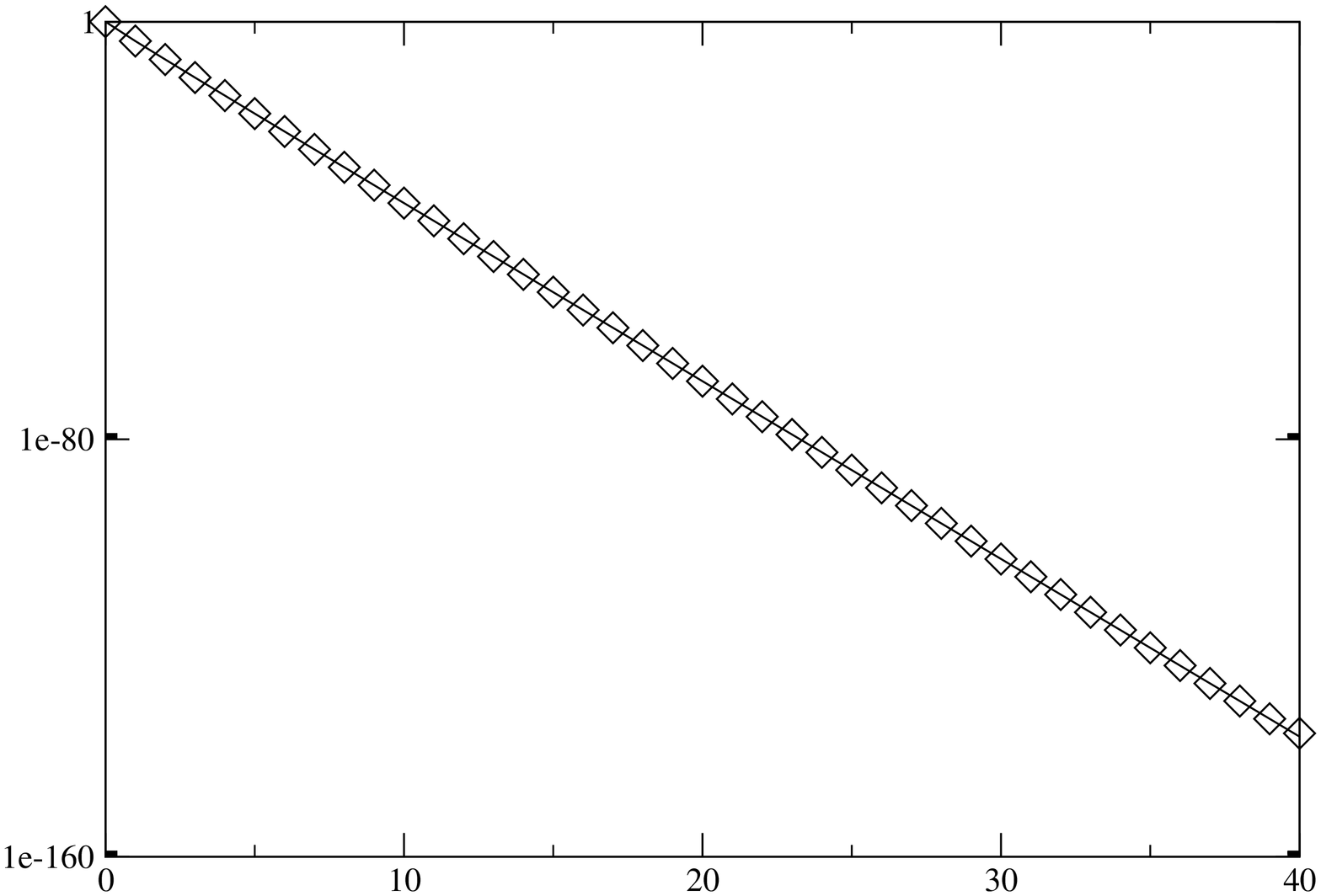}
  \end{center}
  \caption{Correlation function on an $80^2$-lattice at $\beta=0.01$.}
  \label{hundredorders}
 \end{minipage}
\end{figure}

An efficient trick \cite{snakealgorithm}
allows us to measure an exponentially suppressed signal with a linear
effort. The correlation function is exponentially suppressed in the
high-temperature phase and is thus hard to measure at large distances.
The correlation function is the ratio of two partition functions and it can be
rewritten as the product of many fractions which can be measured individually:
\beq\langle\sigma_0^1\sigma_x^1\rangle=\frac{Z(x)}{Z(0)}
=\frac{Z(1)}{Z(0)}\frac{Z(2)}{Z(1)}\dots\frac{Z(x)}{Z(x-1)}.\eeq
In figure \ref{hundredorders} we compare the numerical results at very 
high temperature with the analytical expression \cite{JimboMiwa}.

\section{Conclusion}

Using D-theory we have constructed a new efficient algorithm for the Ising
model. We have an improved estimator for the susceptibility.
Combining the algorithm with the snake algorithm one can measure the correlation
function in regions otherwise hardly accessible. We would
like to apply this approach to other theories, possibly gauge theories.

\end{document}